\documentclass[fleqn,usenatbib]{mnras}

\usepackage{newtxtext,newtxmath}

\usepackage[T1]{fontenc}

\DeclareRobustCommand{\VAN}[3]{#2}
\let\VANthebibliography\thebibliography
\def\thebibliography{\DeclareRobustCommand{\VAN}[3]{##3}\VANthebibliography}

\usepackage{graphicx}	
\usepackage{amsmath}	
\usepackage{amssymb}	

\title[Outbursts of OJ 287]{The April--June 2020 super-outburst of OJ 287 and its long-term multi-wavelength lightcurve with Swift: binary supermassive black hole and jet activity}

\author[S. Komossa et al.]{
S. Komossa,$^{1}$\thanks{E-mail: astrokomossa@gmx.de (SK)}
D. Grupe,$^{2}$
M.L. Parker,$^{3}$
M.J. Valtonen,$^{4,5}$
J.L. Gomez,$^{6}$
A. Gopakumar,$^{7}$
\newauthor
L. Dey$^{7}$
\\
$^{1}$Max-Planck-Institut f\"ur Radioastronomie, Auf dem H{\"u}gel 69, 53121 Bonn, Germany\\
$^{2}$Dept. of Physics, Earth Science, and Space System Engineering, Morehead State University, 235 Martindale Dr, Morehead, KY 40351, USA\\
$^{3}$European Space Agency (ESA), European Space Astronomy Centre (ESAC), E-28691 Villanueva de la Canada, Madrid, Spain \\
$^{4}$ Finnish Centre for Astronomy with ESO, University of Turku, FI-20014, Turku, Finland \\
$^{5}$ Department of Physics and Astronomy, University of Turku, FI-20014, Turku, Finland \\
$^{6}$ Instituto de Astrofísica de Andalucía-CSIC, Glorieta de la Astronomía s/n, E-18008 Granada, Spain, \\
$^{7}$ Department of Astronomy and Astrophysics, Tata Institute of Fundamental Research, Mumbai 400005, India \\
 } 

\date{Accepted 6 July 2020. Received 3 July 2020; in original form 30 May 2020}

\pubyear{2020}

\begin{document}
\label{firstpage}
\pagerange{\pageref{firstpage}--\pageref{lastpage}}
\maketitle

\begin{abstract}
We report detection of a very bright  X-ray-UV-optical outburst of OJ 287 in April--June 2020; the second brightest since the beginning of our Swift multi-year monitoring in late 2015.
It is shown that the outburst is predominantly powered by jet emission. 
Optical-UV-X-rays are closely correlated,
and the low-energy part of the XMM-Newton spectrum displays an exceptionally soft emission component consistent with a synchrotron origin. A much harder X-ray powerlaw component ($\Gamma_{\rm x}=2.4$, still relatively steep when compared to expectations from inverse-Compton models) is detected out to 70 keV by NuSTAR. 
We find evidence for reprocessing around the Fe region, consistent with an absorption line. If confirmed, 
it implies matter in outflow at $\sim$0.1c. 
The multi-year Swift lightcurve shows multiple episodes of flaring or dipping with a total amplitude of variability of a factor of 10 in X-rays, and 15 in the optical--UV.
The 2020 outburst observations are consistent with an after-flare predicted by the binary black hole model of OJ 287, where the disk impact of the secondary black hole triggers time-delayed accretion and jet activity of the primary black hole. 
\end{abstract}



\section{Introduction}



The last few years have seen the first direct detection of high-frequency gravitational waves (GWs) from
merging stellar-mass black holes 
\citep[e.g.][]{Abbott2016, Abbott2019}.
Coalescing supermassive binary black holes
(SMBBHs), formed in galaxy mergers, are the loudest sources of low-frequency GWs in
the universe \citep{Centrella2010}. Therefore, an intense electromagnetic search for wide and close systems in all stages of
their evolution is currently ongoing \citep[review by][]{Komossa2016}. While wide pairs can be identified by spatially-resolved imaging
spectroscopy, we rely on indirect methods of detecting the most compact, evolved systems. 
These are well beyond the ``final parsec'' in their evolution 
\citep{Begelman1980, Colpi2014}, 
in a regime where GW emission contributes to orbital shrinkage.
Semi-periodicity in lightcurves has been a major 
tool for selecting small-separation SMBBH systems. 

OJ 287 is a nearby, bright, and massive blazar at redshift $z=0.306$ \citep{Dickel1967}, and among the best candidates to date
for hosting a compact SMBBH \citep{Sillanpaa1988, Valtonen2016}. 
Its unique optical lightcurve spans more than a century, dating back to 1891. 
It shows pronounced optical double-peaks every $\sim$12 years, which
have been interpreted as arising 
from the orbital motion of a pair of SMBHs, with
an orbital period on that order ($\sim$9 yrs in the system's rest frame). 

While different variants of binary scenarios have been discussed
in the past \citep[e.g.][]{Lehto1996,  Katz1997, Villata1998, 
Liu2002, Britzen2018, Dey2019}, 
the best explored model explains the double peaks as  
the times when the secondary SMBH impacts the disk around the primary twice during its ~12.06 yr
orbit ("impact flares" hereafter). The most recent orbital two-body modelling 
is based on 4.5 order post-Newtonian dynamics and successfully reproduces the
overall long-term lightcurve of OJ 287 until 2019 \citep{Valtonen2016, Dey2018, Laine2020} (and references therein). It requires a compact binary with a semi-major axis of 9300 AU which is subject
to GR precession of $\Phi$=38 deg/orbit, on an eccentric orbit ($\epsilon$=0.7), with a massive primary of $1.8\times10^{10}$
M$_{\odot}$, and a secondary of $1.5\times10^8$ M$_{\odot}$. 
Independent evidence for a massive primary comes from the host galaxy of OJ 287 and other arguments \citep[e.g.][]{Wright1998, Kushwaha2018a, Nilsson2020}.  
We are carrying out a multi-year, multi-frequency monitoring program of OJ 287, in order to search for epochs of outbursts and explore facets of the binary SMBH model \citep[for first results see][]{Komossa2017,  
Komossa2018,
Myserlis2019, Komossa2020a}. 
Independent of the binary's presence, OJ 287 is a nearby bright blazar, and dense multi-frequency monitoring and high-resolution X-ray spectroscopy are powerful diagnostics of jet and accretion physics in blazars.

Here, we present the detection of a bright outburst of OJ 287 in April--June 2020 with the Neil Gehrels Swift observatory \citep[Swift hereafter;][]{Gehrels2004}; even brighter in UV--X-rays than the observed part of the 2015 "centennial" impact flare, and the second brightest in X-rays since the beginning of Swift observations of OJ 287 in 2005. XMM-Newton and NuSTAR X-ray spectroscopy was used in order to understand the nature of this outburst. (The long-term Swift lightcurve will be analyzed further in upcoming work; Paper II hereafter).
We use a cosmology \citep{Wright2006} with 
$H_{\rm 0}$=70 km\,s$^{-1}$\,Mpc$^{-1}$, $\Omega_{\rm M}$=0.3 and $\Omega_{\rm \Lambda}$=0.7 throughout this paper. 

\section{Data analysis and spectral fits}

\begin{table}
\scriptsize
	\centering
	\caption{Summary of X-ray observations (see Sect. 2 for UVOT data).}
	\label{tab:obs-log}
	\begin{tabular}{lcllc}
		\hline
		mission & band (keV) & date ($t_{\rm start}$) & MJD & $\Delta t$ (ksec)\\
		\hline
		Swift XRT & 0.3-10 & 2015 Nov.--2020 June & 57354--59012 & 0.5-2 \\
		XMM-Newton & 0.2-10 & 2020 April 24 & 58963 &  15 \\
		NuSTAR & 3-79 & 2020 May 4 & 58973 & 29 \\
		\hline
	\end{tabular}
\end{table}

\begin{figure}
\includegraphics[clip, trim=1.8cm 5.6cm 1.3cm 2.6cm, width=\columnwidth]{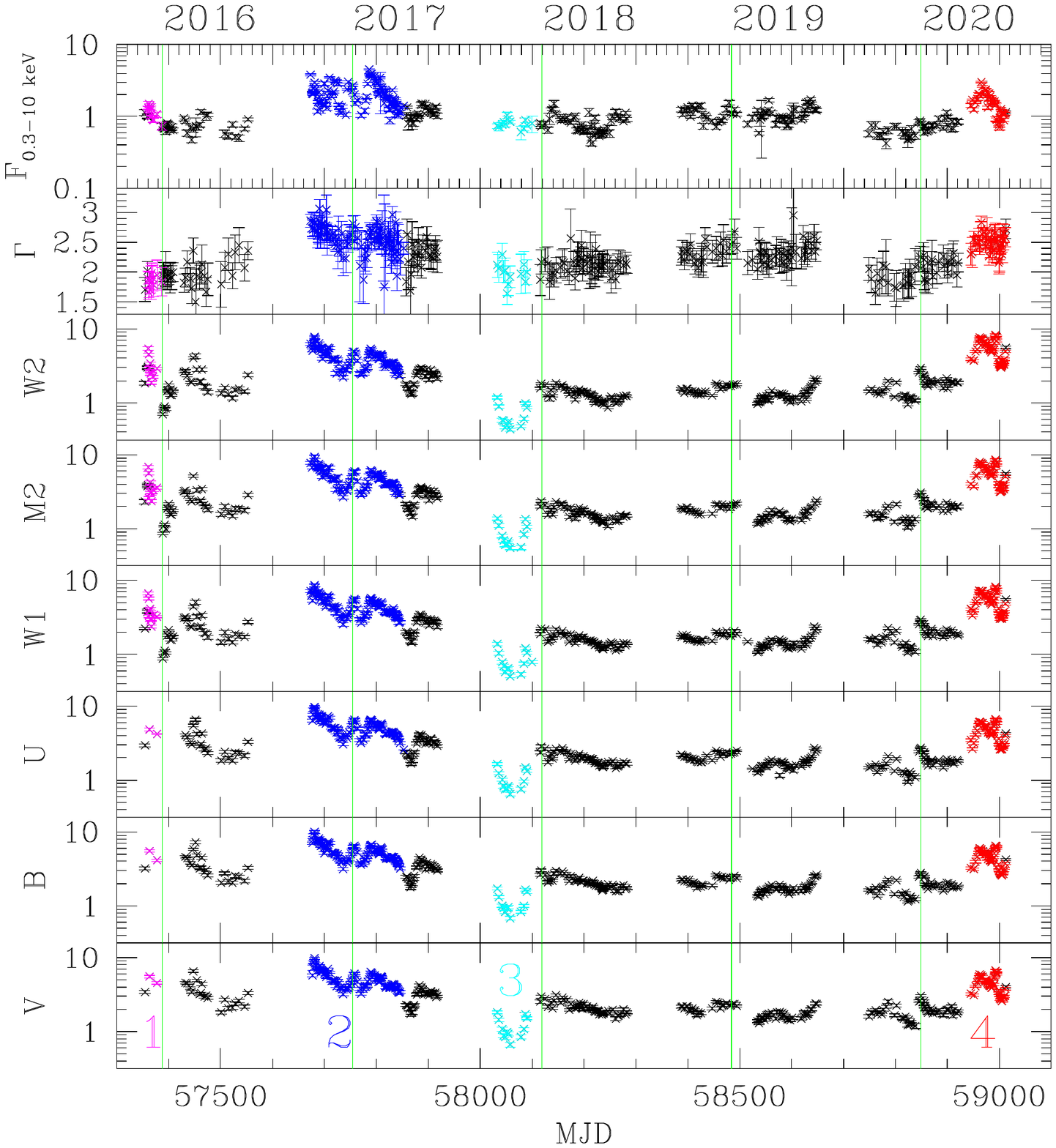}
    \caption{Swift X-ray to optical lightcurve of OJ 287 since the start of our monitoring, between 12/2015 and 6/2020. The observed absorption-corrected X-ray flux (0.3-10 keV) and the extinction-corrected optical-UV fluxes are in 10$^{-11}$ erg/s/cm$^2$. $\Gamma_{\rm x}$ is the X-ray powerlaw photon index. Four epochs of outbursts/low-states are marked in colour: (1) a second peak of the 2015 centennial ``impact flare'', (2) the 2016-2017 outburst, (3) a deep minimum state, and (4) the April-June 2020 outburst. The vertical green lines mark January 1st of each of the years between 2016 and 2020.}
\label{fig:lc-Swift}
\includegraphics[clip, trim=2.0cm 5.6cm 1.3cm 5.2cm, width=\columnwidth]{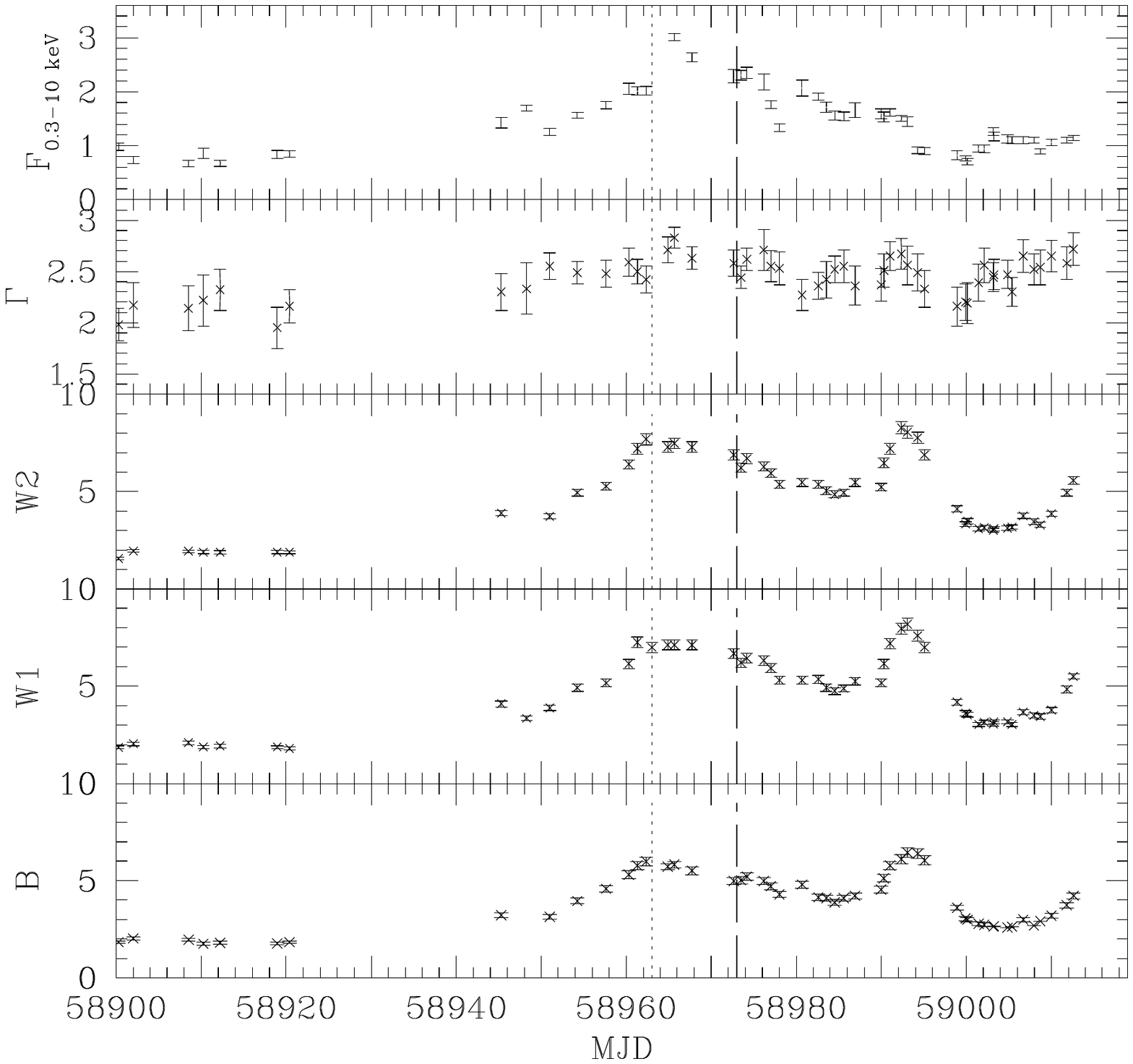}
    \caption{Zoom on the time interval around the 2020 outburst in selected bands (fluxes and units as in Fig. 1). The times of our XMM-Newton (dotted) and NuSTAR (dashed)
    observations are marked by vertical lines. The main outburst peaking in late April (MJD 58962), which is the focus of this Letter, is followed by a shorter second flare which is seen in the optical-UV bands but is less pronounced in X-rays. The last data point is from June 12, 2020. 
    }
    \label{fig:zoom-Swift}
\end{figure}

\begin{table*}
\scriptsize
	\centering
	\caption{Results from XMM-Newton and NuSTAR spectral fits. 
	Absorption was fixed at the  Galactic value $N_{\rm H, Gal}$, except when noted otherwise. 
	Parameters and abbreviations are as follows: (1) Models: pl = powerlaw, bbdy = black body, logpar = logarithmic parabola model; 
	(2) absorption in units of 10$^{20}$ cm$^{-2}$; (3) powerlaw photon index; (4) unabsorbed powerlaw or log-parabola flux from 0.5--10 keV in units of 10$^{-12}$ erg/s/cm$^2$; (5) $kT_{\rm BB}$ in units of keV; (6) bbdy emission in units of $10^{-5}\times(L/10^{39}\mathrm{~erg/s})/[(D/10\mathrm{kpc})(1+z)]^2$, where $L$ and $D$ are the source luminosity and distance; (7) curvature parameter $\beta$ and (8) spectral index $\alpha$ of the log parabola model; (9) goodness of fit $\chi^2_{}$ and number of degrees of freedom. For NuSTAR data, the pl flux is given from 3--50 keV. When no errors are reported, the quantity was fixed. }
	\label{tab:spec-fits}
	\begin{tabular}{lccccccccccc}
		\hline
		model & $N_{\rm H}$ & $\Gamma_1$ & $f_1$ & $\Gamma_2$ & $f_2$ & $kT$ & $f_\mathrm{BB}$ & $\beta$ & $\alpha$ & $f_\mathrm{lp}$ & $\chi{^2}$/$n{_\mathrm{dof}}$ \\
		(1) & (2) & (3) & (4) & (3) & (4) & (5) & (6) & (7) & (8) & (4) & (9) \\
		\hline
		XMM & & & & &  &  & & & & & \\
		\hline
		pl & 2.49 & $2.82\pm0.01$ & $38.5\pm0.1$ & - & - & - & - & -& - & - & 476/284 \\
		pl + pl & 2.49 & $2.84\pm0.01$ &  $38.1\pm0.2$ & $0.0\pm0.4$ & $0.9_{-0.1}^{+0.2}$ &-&-&-&-&-& 380/282 \\
		pl + pl, $N_{\rm H}$ free & $3.4_{-0.8}^{+0.9}$ + 2.49 & $3.1\pm0.1$ & $38.0\pm1.0$ & $1.7_{-0.3}^{+0.2}$ & $5_{-2}^{+3}$ &-&- & & & & 343/281 \\
		pl + bbdy & 2.49 & $2.70\pm0.01$ & $35.2\pm0.3$ & - & - & $0.152\pm0.003$ & $8.0\pm0.7$ &-&-& -& 343/282 \\
		pl + pl + bbdy & 2.49 & $2.76\pm0.03$ & $35.6\pm0.3$ & $0\pm1$ & $0.6_{-0.2}^{+0.4}$ & $0.16\pm0.01$ & $6.0\pm1.0$ &- & &  & 311/280 \\ 
		logpar + pl & 2.49 & 2.2 & $17.3\pm0.6$ & - & - &- &- & $0.8\pm0.1$ & $3.22 \pm0.03$ & $9.1\pm0.3$ & 286/282 \\
		\hline
		NuSTAR &  & & &  & & & & & & & \\
		\hline
		pl & 2.49 & $2.36\pm0.06$ & $6.1\pm0.2$ & - &-& -&-&-&-&-& 74/74 \\
        pl ($>$ 10 keV) & 2.49 & $2.2\pm0.2$ &  &  & &  & & & & &  \\
		\hline
	\end{tabular}
\end{table*}

\subsection{Swift} 
We have monitored OJ 287 since December 2015 (Tab. \ref{tab:obs-log}, Fig. \ref{fig:lc-Swift}, which also includes some Swift data sets from other programs and PIs). The April-May 2020 outburst was typically covered with a cadence of 1-3 days, while the cadence was $\sim$2-10 days at other epochs. Long gaps of several months occur each year when OJ 287 is in Swift sun constraint.  

Most of the time, the Swift X-ray telescope \citep[XRT;][]{Burrows2005} was operating in photon counting mode 
with typical exposure times of 0.5-2 ksec. 
For X-ray analysis, source photons were extracted within a circle of radius 47\arcsec (equivalent to 20 detector pixels). 
The background was determined in a nearby circular region of radius 236\arcsec. X-ray spectra in the band (0.3-10) keV were generated and then analyzed with the software package XSPEC \citep[version 12.10.1f;][]{Arnaud1996}. 

Spectra were fit with single powerlaws of photon index $\Gamma_{\rm X}$ 
adding Galactic absorption with a column density $N_{\rm H, Gal}=2.49\times10^{20}$ cm$^{-2}$. Photon indices range between $\Gamma_{\rm X}$=1.6--3.0 (Fig. \ref{fig:lc-Swift}, \ref{fig:zoom-Swift}), with a general trend of steepening as OJ 287 becomes X-ray brighter. 

We also observed OJ 287 with the UV-optical telescope (UVOT; Roming et al. 2005) and typically in all six filters 
[V (5468\AA), B(4392\AA), U(3465\AA),  UVW1(2600\AA), UVM2(2246\AA), UVW2(1928\AA); where values in brackets are the filter central wavelengths] in order to obtain reliable spectral energy distribution (SED) information of this rapidly varying blazar. 
Observations in each filter were co-added using the task {\em{uvotimsum}}.
Source counts in all six filters were then selected in a circle of radius 
5\arcsec ~and the background was determined in a nearby region of radius 20\arcsec.
The background-corrected counts were then converted into 
fluxes based on the latest calibration as described in \citet{Poole2008} and \citet{Breeveld2010}. 
The UVOT data were corrected for Galactic reddening of $E_{\rm{(B-V)}}$=0.0248 \citep{Schlegel1998}, with a correction factor in each filter according to Equ. (2) of \citet{Roming2009} and based on the reddening curves of \citet{Cardelli1989}. 

\subsection{XMM-Newton}
Our XMM-Newton \citep{Jansen2001} observation of OJ 287 was 
carried out in small window mode for 15 ksec from 2020-04-24 21:13:18 to 2020-04-25 
01:23:18 UTC when OJ 287 was near the maximum of its outburst 
(observation id 0854591201). The effective exposure time was 9 ksec, after removing an epoch of flaring particle background. 

The XMM-Newton data were reduced using the Science Analysis 
Software (SAS) version 18.0.0. EPIC-pn and EPIC-MOS spectra were 
extracted in a circular region of 
20\arcsec centered on the 
source position and background photons were collected in a nearby 
region of $\sim50$\arcsec~for the pn and $\sim100$\arcsec~for the MOS instruments. A lightcurve analysis of the 2020 data did 
not reveal significant short-time variability beyond the 3 sigma level, and therefore the spectra were analyzed as a 
whole without splitting into different flux states.   
Inspection of the RGS spectrum did not reveal significant narrow spectral features, and the
data were not analyzed further. 

\begin{figure}
\includegraphics[width=6.5cm] {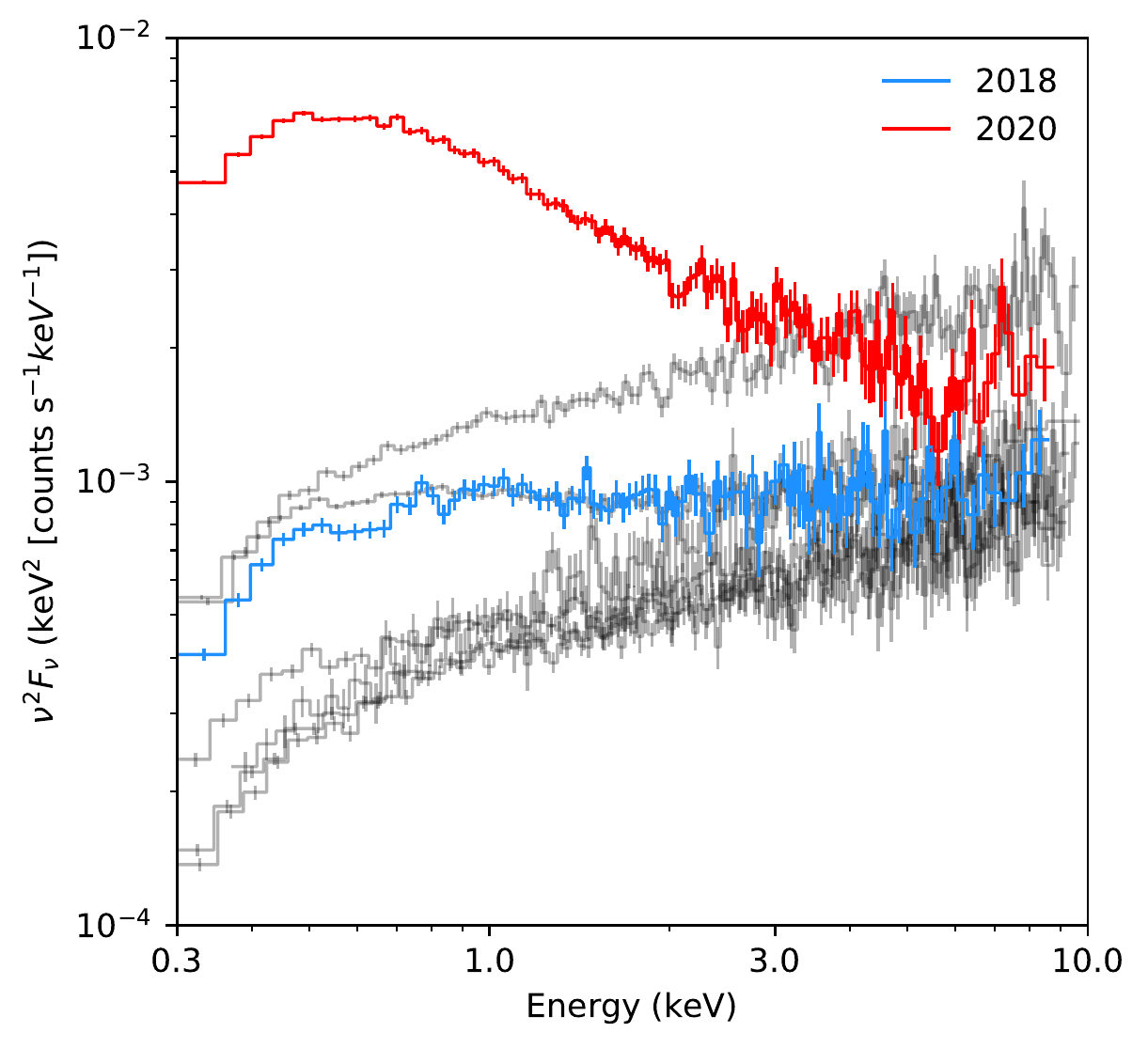}
    \caption{Comparison of all XMM-Newton (EPIC-pn) spectra of OJ 287 between 2005 and 2020 (corrected for effective area of the detector, and without applying any model fits). Our 2020 and 2018 data are highlighted in red and blue, respectively. A strong soft emission component dominates the 2020 spectrum. }
    \label{fig:XMM-allobs}
\end{figure}

\begin{figure}
\includegraphics[width=6.7cm] {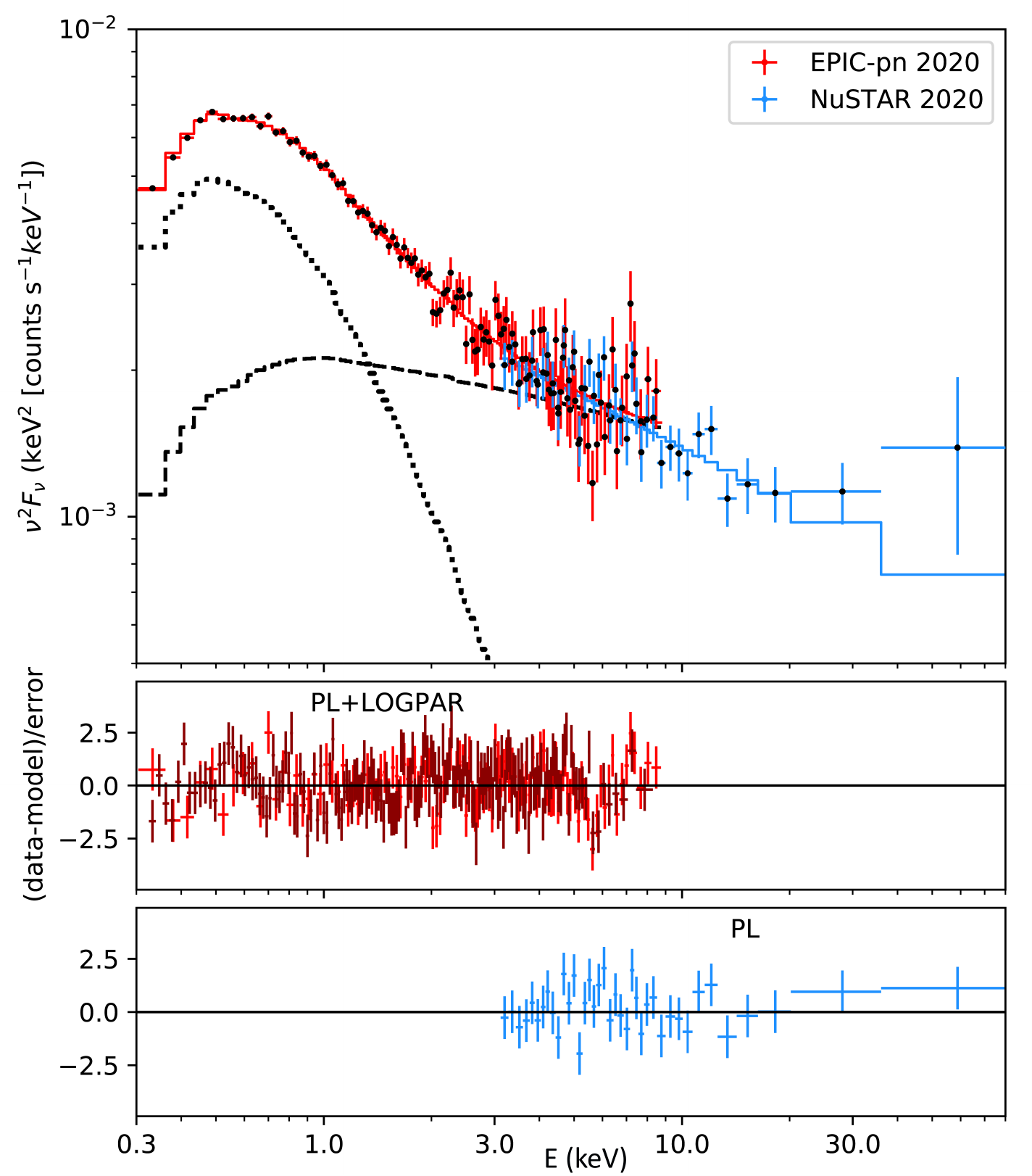}
    \caption{Best-fit log-parabola plus powerlaw model of OJ 287 observed with XMM-Newton, folded with the instrumental response and uncorrected for Galactic absorption. The dotted and dashed lines represent the log-parabola and powerlaw contributions, respectively. The NuSTAR data and fit are added to the plot. The second panel displays the residuals for the best-fit log-parabola model (pn and MOS data), while the third panel displays the residuals of the NuSTAR fit. 
    }
    \label{fig:bestfit2020}
\end{figure}

For further analysis, spectra were binned to a signal-to-noise ratio of at least 6, and to oversample the instrumental resolution by a factor of 3, 
and fit with several emission models (Tab. ~\ref{tab:spec-fits}) 
with absorption fixed to the Galactic value \citep[modeled with 
TBnew;][]{Wilms2000} or left free (at $z=0.3$). OJ 287 is a very bright X-ray source. Fitting is based on $\chi {^2}$ statistics. 
Overall, the spectrum shows a very soft emission component, 
a harder component up to 10 keV, and possible spectral structure in 
the Fe-line region (Fig. \ref{fig:XMM-allobs}, \ref{fig:bestfit2020}). The latter is independently present 
in both, the EPIC-pn and EPIC-MOS data. It is best fit by an 
absorption line of EW = 0.1 keV, at a restframe
energy of $7.45\pm0.05$ keV.
This would correspond to an outflow
with a velocity of 0.067c assuming the line is produced by iron Fe XXVI or 0.1c if it is produced by Fe XXV. Adding a Gaussian absorption line to the best fit model improves the fit by $\Delta\chi^2=17$, for two degrees of freedom, 
which corresponds to a significance of $\sim3.7\sigma$. However, after correcting for the number of trials, assuming 20 resolution elements 
between 7 and 10 keV for the EPIC cameras, the false alarm probability is raised to $4\%$, so the line significance is $\sim2\sigma$. Therefore, its presence has to be confirmed in deeper future observations.  
While single-component broad-band models are unsuccessful, the XMM-Newton spectrum is best fit by the curved log-parabola plus flat powerlaw model with cold absorption 
at the Galactic value (Tab. \ref{tab:spec-fits}). 

For comparison, previous observations of OJ 287 between 2005 and 2018 
were extracted from the XMM-Newton archive (PIs: S. Ciprini \citep{Ciprini2007}, R. Edelson,
S. Komossa/N. Schartel). The EPIC-pn data were reduced similar to our 
2020 observation and were fit with single or double powerlaws. 
During the 2020 observation, OJ 287 was in its highest state so far 
covered by XMM-Newton spectroscopy, and dominated by a strong soft emission component (Fig. \ref{fig:XMM-allobs}).  

\subsection{NuSTAR}

We observed OJ 287 with NuSTAR \citep{Harrison2013}.
The observation (sequence-id 90601616002) was carried out on 2020-05-04 starting at 20:36:09 UTC with an effective exposure time of 29.5 ksec for FPMA and 29.3 ksec for FPMB.
The data were reduced using the latest NuSTAR data analysis software 
(NuSTARDAS) version 1.9.2. 
Source photons were extracted in a circular region of 30\arcsec ~centered on the source position. Background photons were collected 
in a nearby region of radius $\sim100$\arcsec ~on the same chip. 
Spectra were binned 
to a signal-to-noise ratio of at least 6, and to oversample the instrumental resolution by a factor of 3. We fit the spectra separately and we allow for a difference in normalisation and index between the two detectors. As this difference is small ($<1\%$), we 
report only the FPMA values. Source emission is detected out to $\sim$70 keV, with a 2$\sigma$ detection above 40 keV. 
The NuSTAR spectrum is well fit with a single powerlaw model (Tab. 2), without any significant residuals. No strong Fe absorption line is detected. Its presence in both XMM spectra, or its faintness in the later NuSTAR spectrum, therefore either is a statistical fluctuation, or else the feature is short-lived, or Fe becomes completely ionized as the flare continues and thus escapes detection with NuSTAR. 
There is a hint that the NuSTAR spectrum flattens
at high energies, but adding a second powerlaw does not improve the fit and the parameters cannot be constrained.

\section{Results and discussion} 

\subsection{Timing and spectroscopy}

Four epochs stand out in our Swift lightcurve of OJ 287 (Fig. \ref{fig:lc-Swift}): (1) The (late phase of) the December 2015 "centennial flare", interpreted as the disk crossing of the secondary SMBH based on higher-cadence optical data \citep{Valtonen2016, Valtonen2019}. 
(2) The long-lasting 2016-2017 flare which was the brightest in X-rays with Swift and with a soft spectrum \citep{Verrecchia2016, 
Grupe2017, Komossa2017, Kushwaha2018b, Kapanadze2018}. The event was accompanied by a VHE detection \citep{Mukherjee2017}.
(3) A sharp and symmetric deep low-state in late 2017 in all optical \citep{Valtonen2020} and UV bands which is absent in X-rays (to be discussed further in Paper II).
(4) The April 2020 outburst, where OJ 287 reached the second-brightest X-ray state during the Swift monitoring.  
Here, our focus is the 2020 outburst, and X-ray analysis has been done with the following key questions in mind: 

What does the variability imply about the emission site ?  At $1.8\times10^{10}$ M$_{\odot}$, an innermost stable orbit of $R_{\rm ISCO} \sim 3 R_{\rm S}$ corresponds to a minimum {\em restframe} lightcrossing timescale of 6.3 days, which is 
larger than observed. 
Daily changes including a factor 1.7 drop in flux within 2 days during the 2020 outburst (Fig. 2) therefore imply an emission site smaller than the last stable orbit of the primary SMBH of OJ 287, then indicating jet activity.  

Is there any wavelength-dependent delay in the peak time of the flare ? 
UV-optical lightcurves follow each other closely and reach their peak quasi-simultaneously (see paper II for details), implying co-spatial emission and small opacities. X-rays follow substructure in the April flare closely, but the two-week flat plateau does not allow to locate the peak precisely.  

Which mechanism drives the softness of the X-ray spectrum: accretion or jet (synchrotron) activity ?  
The very soft X-ray emission component 
could potentially represent emission associated with the inner accretion disk; either a high-energy tail of the big blue bump or reprocessing/reflection of coronal photons off the inner disk. 
Near the peak of the 2020 flare, the observed powerlaw flux 
corresponds to an {\em isotropic} X-ray luminosity of $10^{45}$ erg/s, which would imply an X-ray Eddington ratio of $L/L_{\rm Edd} = 4\times10^{-4}$ ($8\times10^{-2}$) for a BH of mass $1.8\times10^{10}$ M$_{\odot}$ (10$^8$ M$_{\odot}$) if it was accretion driven.
Given the rapid variability, it then has to be the disk of the secondary BH. %
However, there is no other evidence so far for a long-lasting disk around the secondary, and the quasi-simultaneous variability in all bands from the optical to X-rays strongly argues for a synchrotron origin of the emission. 

Even though blazars often 
show a synchrotron component in the X-ray band \citep{Urry1996, Donato2005}, it is interesting to note that their synchrotron component is rarely as soft
as in OJ 287
(see paper II for further discussion). 
We find that OJ 287 generally exhibits a "softer-when-brigther" variability pattern in our multi-year Swift lightcurve{\footnote{ also seen on long timescales when combining a few Einstein, EXOSAT, ROSAT, and ASCA data (\citet{Isobe2001}; but see \citet{Seta2009})}} -- with the exception of the epoch around the 2015 impact flare when the X-ray spectrum was rather hard.   

In summary, the various observations imply that the April 2020 outburst is not dominated by accretion-disk emission but rather by non-thermal emission from the jet, further corroborated by the Effelsberg detection of a (delayed) radio flare \citep{Komossa2020b}, and by the  detection of high polarization of the optical flare of OJ 287 first reported by \citep{Zola2020}.  

\subsection{Binary black hole model} 
In the context of the binary SMBH model for OJ 287 as reviewed by \citet{Dey2019}, there are several potential sites of UV--X-ray emission, which may become bright at different epochs: 
First, the impact flare (bremsstrahlung) from the secondary, as it impacts the accretion disk around the primary, causing a two-sided expanding bubble \citep{Ivanov1998}. It was last observed in July 2019 \citep{Laine2020} and none is predicted for 2020. 
Second, temporary accretion and perhaps jet emission of the secondary SMBH while and/or after passing the primary's disk \citep{Pihajoki2013}. However, it is unlikely that any secondary SMBH of much lower mass and different spin, and with a temporary disk without large-scale magnetic field, will trigger {\em synchrotron flares} of very similar brightness and spectrum as the primary (see the long-term lightcurve in Fig. 1). 
Third, ``after-flares'' in form of changes in the accretion rate of the primary, after the impact disturbance has travelled to the inner edge of the accretion disk, then later followed by changes in jet activity in response.

\citet{Sundelius1997} \citep[see also][]{Valtonen2009}
predicted the expected after-flares of OJ 287 tidally induced by the secondary.
Based on their model, we expect major {\em accretion} after-flare activity in early January 2020.
Identifying their predicted (accretion) peak in January 2020 with the (jet) outburst reported here in April requires
a time delay of $\sim$4 months 
between {\em accretion disk} and {\em jet} activity, and implies rapid communication between disk and jet.
Factors which determine the actual delay between accretion and jet changes include the disk/corona properties and geometry,
the magnetic field geometry, 
and shock formation in the jet 
\citep[e.g.][]{Marscher2018, Tchekhovskoy2014, Valtonen2019}, which are not yet well understood, and we therefore cannot predict delays from first principles, but can compare with other extragalactic systems where delays were observed.
Overall, the timescale observed in OJ 287 is consistent with the one seen in stellar tidal disruption events \citep{Komossa2016} where accretion flares are typically followed by detectable radio-jet activity within days \citep[e.g.][]{Zauderer2011}, and with the blazar 3C120 for which \citet{Marscher2002} reported a delay of 0.1 yrs between accretion and radio-jet activity.

\section{Summary and conclusions}
We have monitored OJ 287 with Swift since December 2015, revealing multiple epochs of high-amplitude optical--X-ray variability. 
The bright April--June 2020 super-outburst of OJ 287 
has one of the densest quasi-simultaneous optical-UV-X-ray coverages obtained so far for this blazar. We also presented the first XMM-Newton and NuSTAR broad-band X-ray spectroscopy of OJ 287 in outburst.  

Several X-ray spectral features stand out: First, a steep low-energy component ($\Gamma_{\rm x}=2.8$) at peak, rarely that soft in blazars but consistent with a synchrotron origin. Across the flare, OJ 287 is softer when brighter, a pattern also seen in our long-term Swift lightcurve. Second, a powerlaw component detected up to $\sim70$ keV ($\Gamma_{\rm x}\sim2.4$). 
Third, signs of reprocessing in the Fe-line region, which may represent a relativistic outflow if confirmed.   

We find that the outburst is jet-driven and consistent with a binary SMBH model, where the disk impact of the secondary black hole triggers an after-flare in form of time-delayed accretion activity on the primary which is then followed by an increase in jet emission of the primary $\sim$4 months later. 


%



\section*{Acknowledgements}
We would like to thank the Swift, XMM-Newton and NuSTAR teams for carrying out our observations, and our anonymous referee for very useful comments.  

\section*{Data Availability Statement}
Data are available on request. 





\begin{thebibliography}{99}

\bibitem[\protect\citeauthoryear{Abbott et al.}{2016}]
{Abbott2016}
Abbott P.B., et al. 2016, Phys. Rev. Lett. 116, 061102

\bibitem[\protect\citeauthoryear{Abbott et al.}{2019}]{Abbott2019}
Abbott P.B., et al. 2019, ApJL 882, L24

\bibitem[\protect\citeauthoryear{Arnaud}{1996}]{Arnaud1996}
Arnaud K.A., 1996, ASPC 101, 17 

\bibitem[\protect\citeauthoryear{Begelman et al.}{1980}]{Begelman1980}
Begelman M.C., Blandford R.D., Rees M.J. 1980, Nature 287, 307

\bibitem[\protect\citeauthoryear{Breeveld et al.}{2010}]{Breeveld2010}
Breeveld A.A., et al. 2010, MNRAS 406, 1687

\bibitem[\protect\citeauthoryear{Britzen et al.}{2018}]{Britzen2018}
Britzen S., et al. 2018, MNRAS 478, 3199

\bibitem[\protect\citeauthoryear{Burrows et al.}{2005}]{Burrows2005}
Burrows D.N., et al. 2005, SSRv 120, 165

\bibitem[\protect\citeauthoryear{Cardelli et al.}{1989}]{Cardelli1989}
Cardelli J.A., et al. 1989, ApJ 345, 245

\bibitem[\protect\citeauthoryear{Centrella et al.}{2010}]{Centrella2010}
Centrella J., et al. 2010, RevModPhys 82, 3069

\bibitem[\protect\citeauthoryear{Ciprini et al.}{2007}]{Ciprini2007}
Ciprini S., et al. 2007, MmSAI 78, 741

\bibitem[\protect\citeauthoryear{Colpi}{2014}]{Colpi2014}
Colpi M., 2014, SSRv 183, 189 

\bibitem[\protect\citeauthoryear{Dey et al.}{2018}]{Dey2018} 
Dey L., et al. 2018, ApJ 866, 11

\bibitem[\protect\citeauthoryear{Dey et al.}{2019}]{Dey2019}
Dey L., et al. 2019, Universe 5, 108

\bibitem[\protect\citeauthoryear{Dickel et al.}{1967}]{Dickel1967}
Dickel J.R., et al. 1967, AJ 72, 757

\bibitem[\protect\citeauthoryear{Donato et al.}{2005}]{Donato2005}
Donato D., Sambruna R.M., Gliozzi M., 2005, A\&A 433, 1163

\bibitem[\protect\citeauthoryear{Gehrels et al.}{2004}]{Gehrels2004}
Gehrels N., et al. 2004, ApJ 611, 1005

\bibitem[\protect\citeauthoryear{Grupe et al.}{2017}]{Grupe2017}
Grupe D., Komossa S., Falcone A., 2017, Astron. Telegram 10043, 1

\bibitem[\protect\citeauthoryear{Harrison et al.}{2013}]{Harrison2013}
Harrison F.A., et al. 2013, ApJ 770, 103

\bibitem[\protect\citeauthoryear{Isobe et al.}{2001}]{Isobe2001}
Isobe N., et al. 2001, PASJ 53, 79 

\bibitem[\protect\citeauthoryear{Ivanov et al.}{1998}]{Ivanov1998}
Ivanov P.B., Igumenshchev I.V., Novikov I.D., 1998, ApJ 507, 131

\bibitem[\protect\citeauthoryear{Jansen et al.}{2001}]{Jansen2001}
Jansen F., et al. 2001, A\&A 365, L1

\bibitem[\protect\citeauthoryear{Katz}{1997}]{Katz1997}
Katz J.I., 1997, ApJ 478, 527

\bibitem[\protect\citeauthoryear{Kapanadze et al.}{2018}] {Kapanadze2018}
Kapanadze B., et al. 2018, MNRAS 480, 407

\bibitem[\protect\citeauthoryear{Komossa \& Zensus}{2016}]{Komossa2016}
Komossa S., Zensus J.A., 2016, IAUS 312, 13  

\bibitem[\protect\citeauthoryear{Komossa et al.}{2017}]{Komossa2017}
Komossa S., et al. 2017, IAUS 324, 168

\bibitem[\protect\citeauthoryear{Komossa et al.}{2018}]{Komossa2018}
Komossa S., Grupe D., Gomez J.L., 2018, Astron. Telegram 12086, 1

\bibitem[\protect\citeauthoryear{Komossa et al.}{2020a}]{Komossa2020a}
Komossa S., Grupe D., Gomez J.L., 2020a, Astron. Telegram 13658, 1

\bibitem[\protect\citeauthoryear{Komossa et al.}{2020b}]{Komossa2020b}
Komossa S., et al. 2020b, Astron. Telegram 13702, 1

\bibitem[\protect\citeauthoryear{Kushwaha et al.}{2018a}]{Kushwaha2018a}
Kushwaha P., et al. 2018a, MNRAS 473, 1145

\bibitem[\protect\citeauthoryear{Kushwaha et al.}{2018b}]{Kushwaha2018b}
Kushwaha P., et al. 2018b, MNRAS 479, 1672

\bibitem[\protect\citeauthoryear{Laine et al.}{2020}]{Laine2020}
Laine S., et al. 2020, ApJL 894, L1 

\bibitem[\protect\citeauthoryear{Lehto \& Valtonen}{1996}]{Lehto1996}
Lehto H.J., Valtonen M.J., 1996, ApJ 460, 207

\bibitem[\protect\citeauthoryear{Liu \& Wu}{2002}]{Liu2002}
Liu F.K., Wu X.B., 2002, A\&A, 388, L48

\bibitem[\protect\citeauthoryear{Marscher et al.}{2002}]{Marscher2002}
Marscher A.P., et al. 2002, Nature 417, 625  

\bibitem[\protect\citeauthoryear{Marscher et al.}{2018}]{Marscher2018}
Marscher A.P., et al. 2018, ApJ 867, 128

\bibitem[\protect\citeauthoryear{Mukherjee et al.}{2017}]{Mukherjee2017}
Mukherjee R., et al. 2017, Astron. Telegram 10051, 1

\bibitem[\protect\citeauthoryear{Myserlis et al.}{2019}]{Myserlis2019}
Myserlis I., et al. 2019, A\&A 619, 88

\bibitem[\protect\citeauthoryear{Nilsson et al.}{2020}]{Nilsson2020}
Nilsson K., et al. 2020, A\&A, submitted

\bibitem[\protect\citeauthoryear{Pihajoki et al.}{2013}]{Pihajoki2013}
Pihajoki P., et al. 2013, ApJ 764, 5

\bibitem[\protect\citeauthoryear{Poole et al.}{2008}]{Poole2008}
Poole T.S., et al. 2008, MNRAS 391, 1163

\bibitem[\protect\citeauthoryear{Roming et al.}{2005}]{Roming2005}
Roming P.W.A., et al. 2005, SSRv 120, 95

\bibitem[\protect\citeauthoryear{Roming et al.}{2009}]{Roming2009}
Roming P.W.A., et al. 2009, ApJ 690, 163

\bibitem[\protect\citeauthoryear{Schlegel et al.}{1998}]{Schlegel1998}
Schlegel D.J., Finkbeiner D.P., Davis M., 1998, ApJ 500, 525

\bibitem[\protect\citeauthoryear{Seta et al.}{2009}]{Seta2009}
Seta H., et al. 2009, PASJ 61, 1011

\bibitem[\protect\citeauthoryear{Sillanpaa et al.}{1988}]{Sillanpaa1988}
Sillanpaa A., et al. 1988, ApJ 325, 628

\bibitem[\protect\citeauthoryear{Sundelius et al.}{1997}]{Sundelius1997}
Sundelius B., et al. 1997, ApJ 484, 180

\bibitem[\protect\citeauthoryear{Tchekhovskoy et al.}{2014}]{Tchekhovskoy2014}
Tchekhovskoy A., et al. 2014,  MNRAS 437, 2744

\bibitem[\protect\citeauthoryear{Urry et al.}{1996}]{Urry1996}
Urry, M., et al., 1996, ApJ 463, 424

\bibitem[\protect\citeauthoryear{Valtonen et al.}{2009}]{Valtonen2009}
Valtonen M.J., et al. 2009, ApJ 698, 781

\bibitem[\protect\citeauthoryear{Valtonen et al.}{2016}]{Valtonen2016}
Valtonen M.J., et al. 2016, ApJL 819, L37

\bibitem[\protect\citeauthoryear{Valtonen et al.}{2019}]{Valtonen2019}
Valtonen M.J., et al. 2019, ApJ 882, 88

\bibitem[\protect\citeauthoryear{Valtonen et al.}{2020}]{Valtonen2020}
Valtonen M.J., et al. 2020, submitted

\bibitem[\protect\citeauthoryear{Verrecchia et al.}{2016}]{Verrecchia2016}
Verrecchia F., et al. 2016, Astron. Telegram 9709, 1

\bibitem[\protect\citeauthoryear{Villata et al.}{1998}]{Villata1998}
Villata M., et al. 1998, MNRAS 293, L13


\bibitem[\protect\citeauthoryear{Wilms et al.}{2000}]{Wilms2000}
Wilms J., Allen A., McCray R., 2000, ApJ 542, 914

\bibitem[\protect\citeauthoryear{Wright}{2006}]{Wright2006}
Wright E.L., 2006, PASP 118, 1711

\bibitem[\protect\citeauthoryear{Wright et al.}{1998}]{Wright1998}
Wright S.C., et al. 1998, MNRAS 296, 961

\bibitem[\protect\citeauthoryear{Zauderer et al.}{2011}]{Zauderer2011}
Zauderer B.A., et al. 2011, Nature 476, 425

\bibitem[\protect\citeauthoryear{Zola et al.}{2020}]{Zola2020}
Zola S., et al. 2020, Astron. Telegram 13637, 1

\end{thebibliography}




\appendix

\bsp	
\label{lastpage}
\end{document}